**DEGAS 2 model validation study: comparison of measured and modeled helium and deuterium line emission arising from an external gas puff on Alcator C-Mod**

S. G. Baek, J. L. Terry, D. P. Stotler*, B. Labombard, D. Brunner

Plasma Science and Fusion Center, MIT, Cambridge, MA 02139 USA
*Princeton Plasma Physics Laboratory, Princeton, NJ 08540 USA

E-mail: sgbaek@psfc.mit.edu

Abstract: The ability to accurately model and predict neutral transport in the boundary plasma is important for tokamak operation. Nevertheless, validation of neutral transport models can be challenging due to the difficulty in measuring neutral particle distributions. Taking advantage of the localized neutral gas puff associated with the Gas Puff Imaging (GPI) diagnostic on the Alcator C-Mod, a validation study of the neutral transport code DEGAS 2 has been performed for helium and deuterium neutrals. Absolutely calibrated measurements of helium and hydrogen line emission are compared with simulated emission from DEGAS 2, accounting for the measured gas flow rates and employing a realistic geometry. Good agreement in peak brightness and profile shape is found for a deuterium puff case. However, helium line emission measurements are found to be lower by a factor of three than that predicted in the steady state DEGAS 2 simulations for a wide range in plasma density. Discrepancies in the light emission profile shape are evident as well. As possible causes for this discrepancy, two effects are discussed. First is the effect of local cooling due to gas puff. Second is the role of time-dependent turbulence effects on neutral penetration.

1. **Introduction**

Neutrals in tokamak plasmas play an important role in determining particle, momentum, and energy balance [1]. For example, a recent modeling study [2] identifies neutrals as one of the key components in establishing the stable detached divertor operation [3], envisioned for next generation tokamaks for an efficient heat exhaust. While *modeling* neutral transport has been an active research area, *validating* neutral transport models has not been straightforward due to the difficulties in characterizing the neutral particle distribution. The experimental characterization of neutral transport generally relies on measuring the line emission from the neutrals in question. In that case, the study of neutral transport is tightly coupled to accurate modeling of the atomic physics responsible for the line emission [4-6]. Characterization of neutral transport of the "working" gas for the plasma is made further difficult by the imprecise knowledge of the source (e.g., recycling), and by the fact that the source and emission can vary greatly spatially, whereas the diagnostic views are typically limited in spatial coverage and resolution.

Gas Puff Imaging (GPI) experiments [7-11], on the other hand, can avoid most of these difficulties because spatially-localized neutrals are puffed into the plasma at known rates with known source geometries, and line emission from the puffed neutrals is viewed with excellent spatial coverage and resolution. Because of a typical view location at the midplane, modeling plasma parameters is also relatively straightforward as compared to that in the divertor region. GPI is widely used to study edge and scrape-off layer (SOL) turbulence since fluctuations in neutral line emission result from local plasma fluctuations in electron density and temperature that are responsible for regulating plasma particle and heat transport in a tokamak. With GPI, one obtains a time series of cross-field 2D line emission measurements by minimizing the toroidal extent of gas puff in the toroidal ($\varphi$) direction with the alignment of the detector line-of-sight to the background magnetic field line. Although the fluctuation in the measured emission is the quantity of interest for GPI, one can extract important information about the spatial distribution of the neutral particles from the time-averaged emission profile, if the time-averaged profiles of the plasma temperature and density are known and if the emission modeling is sound. For example, one could conclude that the neutral particle distribution obtained from the modeling represents the actual distribution, when the synthetic emission profile matches to the measured emission profile.

The purpose of this work is to describe tests of the modeling capability of the DEGAS 2 neutral transport code [12] for both helium (He) and deuterium molecule ($D_2$) neutral sources, using the measurements from the GPI diagnostics of Alcator C-Mod [13] plasmas. It is worth stressing that one can still only test the *combination* of the transport (e.g., the change in particle trajectories due to scattering) and the atomics physics modeling (e.g., excitation, ionization, and emission). This is because the transport and atomic physics reaction rates are coupled to each other. For example, an ionization process can limit how deep neutral particles penetrate into the hot plasma core. Once the particle spatial distribution is determined from the Monte Carlo particle tracking process in DEGAS 2, the atomic physics model is again used to obtain the emission profile.

The C-Mod GPI validation study presented in this paper extends the previous series of DEGAS 2 validation studies in the following ways: first, we compare the absolute brightness measurements with the steady-state DEGAS 2 simulation results with no free parameters in the modeling. The gas puff rate is evaluated by reconstructing the time history of the absolute gas-flow rate. The optical system and detector responses have absolutely been calibrated for both $D_\alpha$ and He I 587.6 nm line emissions. Second, the availability of the $D_2$ and helium puffs at the same puff location (i.e., the outer midplane) allows studying neutrals that have completely different atomic/molecular physics while, if not completely eliminating, minimizing plasma-wall interactions. Whereas deuterium enters the edge of the plasma as a (diatomic) molecule and undergoes a host of atomic processes [11], helium is an inert and monoatomic gas. Third, the use of helium neutrals on C-Mod guarantees that the measured line emission is from the gas puff only with no intrinsic background line emission because the main gas species for these C-Mod discharges is deuterium. Lastly, the brightness measurements have been conducted over a wide density range, spanning line-average electron densities ($\bar{n}_e$) from $0.84 \times 10^{20}$ to $2.2 \times 10^{20}$ m$^{-3}$ in L-mode plasmas. This range corresponds to Greenwald fractions ($\bar{n}_e/n_G$) from 0.2 to 0.7, where $\bar{n}_e$ is the electron line-averaged density in units of $1 \times 10^{20}$ m$^{-3}$, and $n_G = I_p/(\pi a^2)$ is the Greenwald density where $I_p$ is the plasma current in MA and $a$ is the tokamak minor radius in meters. For the $D_2$ puff case, the edge/Scrape-Off Layer (SOL) plasma density and temperature parameters happens to lie in a similar parameter space to those of the NSTX study [14], although the C-Mod experiments are performed in L-mode plasmas. For the helium-puff cases, the measurements are conducted at $\bar{n}_e/n_G = 0.2$, 0.5, and 0.7. As the Greenwald fraction increases above $\bar{n}_e/n_G = 0.2$, blobby transport [15] dominates particle transport in the far-SOL, and this is known to affect neutral transport [16,17,18]. This wide range provides the stringent test cases for the combined modeling of neutral transport and neutral line emission.

The goal of this paper is to validate the DEGAS 2 code with the experiments conducted on Alcator C-Mod. The absolute brightness and spatial profile of the $D_\alpha$ (656 nm) and helium I (587 nm) line emissions are compared between the experiment and the modeling. A focus is given to the helium puff case because no absolute comparison of the helium line emission has been reported, while good agreement in terms of the profile shape has previously been reported [5]. This paper does not deal with additional development of the DEGAS 2 code. Rather, we detail the experimental calibration, measurement, and comparison processes, running DEGAS 2 with the measured plasma $n_e$ and $T_e$ profiles. We will show that the measured helium line emission is lower than that predicted in DEGAS 2 by a factor of roughly three, and, moreover, the profile shapes do not agree very well over a range of plasma densities. This was somewhat unexpected because a helium atom undergoes only a few interactions in plasma. Local cooling due to gas puff and the impact of plasma fluctuations are qualitatively discussed as a possible cause of this discrepancy. Meanwhile, relatively good agreement is found in the $D_2$ puff case for both the absolute peak brightness and profile shape. This agreement is consistent with a previous GPI comparison with DEGAS 2 using NSTX H-mode plasmas [14], which showed that both the profile shape and the photons per puffed molecule are in good agreement. In that study, however, the peak brightness could not be compared since the temporal variation of the gas flow rate was unknown.

The organization of this paper is as follows. Section 2 outlines the GPI experimental setup with the calibration processes of the flow rate and detector response. Section 3 discusses the DEGAS 2 modeling procedure. Section 4 compares the experimental brightness profiles with the synthetic line emission profiles predicted from DEGAS 2. Section 5 is a discussion of the observed discrepancies between the experiment and simulation. Section 6 summarizes the paper's findings.

## 2. Experimental Setup and Calibration Procedures

The GPI system is one of the key diagnostics for studying edge turbulence on the Alcator C-Mod tokamak. The major radius of the C-Mod tokamak is 0.68 m, and the minor radius is 0.22 m. Table 1 summarizes plasma parameters studied in this paper. The toroidal magnetic field was 5.4 T on axis. The plasma current was 0.5 MA for the plasmas with helium puffs and 1.2 MA with $D_2$ puffs. The line-averaged density ($\bar{n}_e$) ranged from $0.8 \times 10^{20}$ m$^{-3}$ to $2.3 \times 10^{20}$ m$^{-3}$. The experiments in this paper were conducted in two consecutive days.

The GPI system used in this study consists of an optical system that relays the visible light emissions from the gas puff to detectors that record the 2D images of line emissions [19]. It utilizes a gas puff system (described below) along with a nozzle located at the outer midplane, as shown in Figure 1(a). The nozzle is installed in a "split" between the two poloidally displaced sections of an outboard limiter. While not indicated in the figure, the nozzle's plasma-facing side is in the shadow of two other outboard limiters that are toroidally displaced by 56° and 155° respectively. The nozzle is made up of four capillaries, which are vertically distributed (Z = -4.19, -3.39, -2.59, and -1.9 cm) at the same toroidal angle and major radial location (R = 92.66 cm). The nozzle is 1.2 cm (2.1 cm) into the SOL from the separatrix for the He-puffed ($D_2$-puffed) discharges considered here. The neutral helium atoms or deuterium molecules are supplied from the Neutral gas INJection Array (NINJA) system that is a gas storage and delivery system [20]. A pneumatically controlled valve at the storage plenum feeds a 3 m long, 1 mm diameter capillary that ends at the nozzle. As a result it takes about ~130 ms for the gas to reach the nozzle after the valve is opened, and ~1.5 seconds for all the gas to leave the capillary after the valve is closed.

The optical system and detectors consist of an in-vessel telescope, optical fiber cables, and an array of avalanche photodiodes (APDs). Lens-based optics mounted on the outer wall of the vacuum chamber collect the emission along nearly horizontal chords that are approximately normal to the radial-poloidal target plane in front of the nozzle (the red rectangle in Figure 1(a)). At the image plane of the optics is a 9x10 array of optical fibers. The field-of-view (f-o-v) of the array in the target plane is 3.8 cm × 4.4 cm, with each fiber viewing a ~3.8 mm diameter spot in this plane. The distance between the front-end optics and the target plane is 60 cm. The lines-of-sight make a ~5° (10°) angle with the local magnetic field for the He-puffed ($D_2$-puffed) shots. The fibers are brought through the vacuum interface to the APD arrays, where the light is filtered for the desired line emission by an appropriate interference filter as it is coupled to the APD detectors. The voltage signals measured by the APDs are digitized at a sampling rate of 2 MHz.

### 2.1. Flow Rate Calibrations

In this comparison of experiment and modeling there are no free parameters used in the modeling except plasma parameters such as $n_e$ & $T_e$. The absolute line brightness are compared after specifying in the simulation the measured electron temperature and density profiles of plasma (Section 3) and specifying the absolute flow rate of puffed gas at the GPI nozzle. We discuss the measurement of the flow rate in this section.

Since there is no flow rate monitor at the nozzle, the time history of the flow rate is reconstructed by combining the waveforms (1) from the measured APD signal for the fast rising part, and (2) from the time rate of change of the vacuum chamber pressure for the slowly decaying part. Note that the slow decay results from the gas bleeding out of the 3.48 m capillary. The flow rate for the slowly decaying part is calibrated "off-line", i.e. not during the plasma discharge, by reproducing the settings of the NINJA system and puffing into the evacuated vacuum chamber. Here, the NINJA settings refer to the pressure of the gas reservoir and the valve "open" duration time. For example, Figure 2 shows the smoothed helium flow rate waveforms at different NINJA reservoir pressures. These waveforms are derived by taking time derivative of the measured C-Mod chamber pressure, and are representative of the conductance-limited flow rate away from the nozzle. Therefore, the initial part of this flow rate waveform does not represent the actual rise of the flow rate at the nozzle.

The fast rise of the flow rate at the nozzle is assumed to be proportional to the initial rise of the GPI brightness signal, which exhibits a rapid rise to its peak value, as shown in Figure 3(a). The waveform is low-pass filtered to remove the effect of plasma fluctuation in the measured signal. The fast rise and slow decay waveforms are then joined at a time when both waveforms show the similar decay rate (i.e., usually when ~100-200 ms after the start of the injection) in order to reconstruct the flow rate waveform at the nozzle. This reconstructed waveform for the flow rate is then normalized so that the total number of particles that left the reservoir is equal to time integral of the flow rate. The total amount of neutrals puffed into the chamber during the actual plasma discharge is found by measuring the absolute pressure change of the NINJA reservoir using a Baratron type pressure gauge. The gauge is not sensitive to the gas species, so there is no need to apply a gas correction factor.

Figure 3(b) shows an example of this procedure. The waveform in blue is from the red trace in Figure 3(a), and proportional to the He I line emission during the plasma discharge for a view passing in front of the nozzle. The measurement duration is limited by the high-resolution digitizer capability from t = 1.1 s to t = 1.5 s. The trace in black is from the flow rate waveform derived from the chamber pressure change after puffing the neutral gas into C-Mod with no plasma at the same NINJA settings to those used for the actual experiment. The chamber pressure is recorded long enough to capture the complete decay of the flow rate (~15 seconds). Note that this waveform obtained "off-line" is translated in time to match the valve trigger time in the experiment at t = 1.1 second. The GPI signal is then scaled to match the level of the flow rate measured "off-line" to identify the time point where both waveforms start to have a similar decay rate. For example, in Figure 3(b), the decay rates of the blue and black waveforms are essentially the same from t = 1.4 s to t = 1.5 s, consistent with the assumption made for the waveform reconstruction. Therefore, in this particular case, the blue and black waveforms have been joined at t=1.5 s.

Once the shape of the flow rate waveform at the nozzle is reconstructed, the absolute flow rate is determined by normalizing the waveform such that the time integral of the reconstructed waveform in the red trace is equal to the total number of particles that left the NINJA reservoir. In this particular case, the flow rate at the time of interest, i.e., at which the density and temperature profile measurements are available (at 1.3 s), is found to be 467 Pa-l/second, or equivalently $1.11 \times 10^{20}$ particles/second. In deriving the particle number, the neutral particle temperature of 300 K is assumed. Table 2 summarizes the flow rates evaluated in this manner for the four plasma discharges and gas puffs studied in this paper.

For the $D_2$ puff case, a similar procedure is performed to evaluate the flow rate of $D_2$ molecules, except that one needs to account for and subtract the intrinsic "background" $D_\alpha$ emission in the measured APD signals. This is required since the DEGAS 2 model in Sec. 4 simulates only the $D_2$ gas puff. The time history of intrinsic $D_\alpha$ emission is assumed to be that measured in a sister discharge that was a repeat of the one with the $D_2$ puff (but with an 8% higher electron density), but with no puff. The $D_\alpha$ brightness time histories for one of the GPI views from the two discharges are shown in Figure 4(a). As shown in the light-blue trace, the intrinsic emission decreases with the programmed decrease in plasma density after 1.33 second (and plasma current after 1.4 second). This change of the plasma parameter was also the case for the plasma discharge of interest. The intrinsic emission in this second discharge is multiplied by ~0.75 in order to match the pre-puff $D_\alpha$ emission from 1.20 s to 1.23 s. Shown in blue is the $D_\alpha$ signal with the intrinsic contribution subtracted, i.e. the emission signal from the gas puff alone. This allows the reconstruction of the $D_2$ puff flow rate in the manner used for the helium puff. The reconstructed flow rate for the $D_2$ puff analysis is shown in Figure 4(b). For a similar puff rate, the flow rate in the $D_2$ puff case is found to be about 30% higher than that in the low density helium puff case. This nonlinearity between the gas pressure and the flow rate has previously been observed in the NINJA system.

In order to validate the reconstructed flow-rate waveform, 2D axisymmetric time-dependent numerical modeling of a helium gas flow is conducted using ANSYS FLUENT with a simplified geometry. It is assumed that a capillary is a 3.48 meter long straight pipe with a diameter of 1 mm with a constant inlet pressure corresponding to the measured plenum pressure in the NINJA system (e.g., 51392 Pa for shot no. 1160616009). The volume of the modeled capillary was 2.73 $cm^3$. While the outlet pressure was set to 0 Pa initially, FLUENT extrapolates the pressure from the flow in the interior region when the flow becomes supersonic, which is the case near the outlet side in our model. In the initialization process, a low pressure (100 Pa) with a constant temperature of 300 K was assigned in the entire capillary domain (1 mm < x < 3.48 m), except the small region near the inlet side (x < 1 mm). A finite amount of pressure was necessary because FLUET solves in the fluid limit. In our model, the constant inlet pressure was applied for 110 msec, while the experimental valve "open" duration time was set to 90 msec for the shot no. 1160616009. This was to account for a possible delay in the valve response because the main pneumatic valve that separates the plenum side and the vacuum side (capillary + nozzle) is controlled by another valve with a delay time of ~10 msec. After t = 110 msec, an additional "drain volume" of 3.4 $cm^3$ is added in order to take into account of the extra volume from a hand valve, feedthrough tubing, and the fast pneumatic valve in the NINJA system. Since this drain volume is located at the inlet side of the capillary, it is lumped as a single cylinder that is pressurized to the inlet pressure. The model was then continued to capture the long decay part of the flow rate wave form.

Figure xx compares the experimentally reconstructed flow rate waveform with four different modeling results. Shown in purple (cyan) color is the modeled waveforms based on the inlet pressure of the experimentally measured value (51329 Pa) with the laminar (turbulent k-epsilon) viscous model. The flow velocity magnitude profile along the axis shows that the flow speed remains below 50 m/sec except near the last 0.2 m region where the flow becomes choked (~ 1.1 km/sec ). The Reynolds number is in the range of 200 ~ 510, implying the flow is in the laminar regime. With the laminar model, the modeled peak flow rate is found to be about 18% less than the experimentally reconstructed peak flow rate. On the other hand, it is also found that the total mass within the simulation domain (the total drain volume) is about 13% less than that found in our calibration process described above (4.15x$10^{-7}$ kg versus 4.75x$10^{-7}$ kg). To compensate this mass difference, we repeated the modeling with the increase in the inlet pressure by 13% to 58002 Pa. Shown in green (red) color in Figure xx is the modeled waveform based on the increased inlet pressure with the laminar (turbulent k-epsilon) viscous model. In this case, FLUENT can reasonably reproduced the peak flow rate and the total mass injected to the machine. The modeled decay-rate is faster than what is reconstructed from the "off-line" calibration procedure.

### 2.2. Absolute Brightness Calibration of the Optical System and Detectors

The sensitivity of the optical system, including fiber transmission, interference filter transmission, and the detectors was calibrated in order to compare the absolute line emission with predictions made with the DEGAS 2 simulating the gas puff and plasma of interest. This was accomplished using an absolutely calibrated light source, a commercially available Labsphere integrating sphere [21]. The absolute brightness of this continuum source is supplied by the vendor with a nominal accuracy of ~3%, traceable to the NIST [22]. Before the start of C-Mod's 2016 run-campaign, the integrating sphere was taken in-vessel and placed in the front of the optical system. Because the source is continuum emission, and the system sensitivity at the line wavelength of interest (656 nm or 587 nm) is desired, the bandpass of the respective interference filter is characterized. The full width at half maximum (FWHM) of helium I line ($D_\alpha$) filter is 11.45 nm (11.68 nm). The filter band pass is approximately rectangular. Then, the inverse-sensitivity, *S*, for a given view is calculated as: $S = B(\lambda) \times FWHM / V$, where $B(\lambda)$ is the brightness of the integrating sphere at the wavelength of interest, and *V* is the voltage measured when viewing the sphere. In the calibration process, the APD detector high-voltages are set to those values used in the plasma experiment. Once the inverse-sensitivity, *S,* is characterized, the voltage signals measured during the experiments are multiplied by *S* to evaluate the absolute line brightness for each view.

At the end of the 2016 run campaign, the optical system calibration was repeated because the in-vessel optics can degrade during plasma operation over the course of the run campaign. This can be caused by browning of the refracting optics and/or mirror coating due to boronization. The sensitivity degradation of 62% (69%) was in fact observed at the He I line ($D_\alpha$) wavelength, as compared to the pre-campaign calibration. In order to account for this sensitivity change, the linear decrease in the sensitivity with the number of plasma shots is assumed. This linear interpolation provides a correction factor (C.F.) to be used in the inverse-sensitivity: $S = S_0 \cdot$ C.F., where $S_0$ is the inverse-sensitivity measured prior to the 2016 campaign, and C.F. is 1.22 (1.17) for the He I line ($D_\alpha$) for the corresponding experimental run day.

### 3. Predictions from the DEGAS 2 Simulations

DEGAS 2 is a Monte Carlo neutral transport code, which allows studying of neutral particle distributions in both real space and phase space. An atomic physics model is used to describe the interactions of neutrals with the plasma. Ionizations and radiative emissions are treated by a collisional radiative model, which assumes that higher-order excited states decay to the ground state in a much faster time scale than that of the change in the ground state population. For example, the helium collision radiation model by Goto [11,23] is used in DEGAS 2. Even though the model neglects two meta-stable $n = 2$ states ($2\ ^1S$, and $2\ ^3S$) in helium energy levels, this so-called $N_p = 1$ model is previously shown to be valid in the density and temperature space of interest for a typical time scale on the order of microseconds [5]. Singlet-triplet wave-function mixing [23] is also found to have a negligible effect on the effective reaction rates at a typical C-Mod magnetic field at the low-field-side edge (4.1 T).

Three physical processes are considered in modeling helium neutral transport in DEGAS 2: (1) electron impact ionization, (2) line emission rate, and (3) neutral-ion scattering. The first two come from the Goto model. The steady-state neutral particle distributions are found in the 3D space by tracking 10 million particles. The effective volumetric photon emission rate at 587.7 nm, $P_{5877} = E_{23} A_{23} (n_{3^3D}/n_{1^1S})/n_e$, is pre-tabulated as a function of the electron density and temperature in units of (eV cm$^3$ /sec), where $n_{3^3D}$ is the density at the $3^3D$ state, and $A_{23}$ is the spontaneous radiative transition rate: $A_{23} = A_{3\ ^3D \to 2\ ^3P} = 7.07 \times 10^{-7}$ sec$^{-1}$. $E_{23} = 2.11$ eV is the transition energy from the 3 $^3$D state to the 2 $^3$P state, as compared to the helium first ionization energy of 24.6 eV. The Goto model shows that emissions are suppressed at temperatures below 10 eV corresponding to the far SOL. Further, the emission rate coefficient decreases by an order of magnitude for $n_e > 10^{20}$ m$^{-3}$ relative to that at $n_e = 10^{18} \sim 10^{19}$ m$^{-3}$. Thus, the emission profile is expected to be peaked around the separatrix. The volumetric emission power is calculated by $P_{vol} = P_{5877}(n_e, T_e) \times n_{1^1S} \times n_e$ where $n_{1^1S}$ is the neutral density at the ground state and $n_e$ is the electron density. On the other hand, deuterium molecules undergo a series of atomic/molecular interactions such as scattering, ionization, dissociation, and recombination. As a result, the $D_\alpha$ emission comes from not only the deuterium atom but also from excited atoms created by molecular dissociation. A complete series of the atomic processes in deuterium molecules used in DEGAS 2 is discussed in more detail in elsewhere [4, 11].

In addition to the absolute flow rate, other critical inputs to DEGAS 2 are magnetic equilibria and kinetic profiles to map the plasma density and temperature onto the 3D space. In the toroidal direction, a toroidal symmetric geometry from $\varphi = -80$ deg. to $\varphi = 30$ deg. is constructed with the nozzle located at $\varphi = -18.51$ deg. The toroidal extent of the nozzle section is 0.06 deg. or an arc length of 0.97 mm. To simulate the actual nozzle's four capillary outputs, four corresponding rectangular sources of height of 1 mm are modelled at the nozzle locations with a gap of 0.7 mm between them. In the code, the particle injection angular distribution at the nozzle has a cosine dependence in the angular direction from $\theta = -\pi/2$ rad. to $\theta = +\pi/2$ rad. with respect to the capillary output surface. Particle temperature is assumed to be 300 *K*. An EFIT code is used to generate magnetic equilibrium files for the shots of interest. The plasma domain in the R-Z space spans from R= 0.8 m to R = 0.9265 m, and from Z = -0.2 m to Z = 0.15 m. In this 2D space, there are 10~30 contours allocated inside the separatrix, and 20~30 contours used outside the separatrix. In between the contours, both the density and temperature are constant.

As shown in Figure 5, the density and temperature profiles are measured with the Thomson and Mirror Langumir Probe (MLP) systems. The MLP system is ~ 12 cm above the GPI view location. Profiles measured with the edge Thomson system are smoothly joined to the SOL profiles measured with MLP. The ion temperature (density) is assumed to be the same as the electron temperature (density). The electron temperature at the separatrix varies from 10 eV to 60 eV depending on the line-averaged density. It is worth mentioning the similarity in the profiles between the plasma discharge with the $D_2$ puff (shot no. 1160617031) and the plasma discharge with helium puff at the lowest density (shot no.1160616009). Although both the line-averaged density and plasma current differ by nearly a factor of two, it is known that the SOL density profile is a function of the Greenwald density fraction. This profile similarity provided additional confidence in the measured SOL profiles. The minimum $T_e$ is assigned to be 1 eV, and the minimum $n_e$ is assigned to be $10^{17}$ m$^{-3}$.

Figures 6(a) and (b) shows the examples of the DEGAS 2 simulation results. Figure 6(a) shows the 2D neutral helium density distribution at the toroidal location in front of the nozzle. The neutral density decreases exponentially from ~$10^{21}$ #/m$^3$ to ~$10^{15}$ #/m$^3$ inside the separatrix due to a strong ionization effect as plasma temperature increases above ~30 eV. The black line denotes the EFIT separatrix. The volumetric photon emission rate is shown in Figure 6(b) with the overlaid field-of-view of the detector in black. Since the photon emission is dependent not only on neutral density but also on electron density, the photon emission profile is found to be peaked just outside the separatrix, unlike the neutral density profile. Based on these outputs, the DEGAS 2 code internally constructs the line-integrated emission along the specified sightlines from the front-end optics system to the target plane and beyond to simulate the camera images.

In comparing these outputs to the experimental measurements, the radial alignments are applied to the input $n_e$ and $T_e$ profiles to compensate for the discrepancy found in the radial mapping between the GPI and MLP measurements. While EFIT is generally used to identify the flux coordinates based on poloidal and toroidal fluxes, an additional physical constraint is needed to align the two independent measurements due to the uncertainties in EFIT mapping. In this paper, we force that the change in the turbulence phase velocity from the electron diamagnetic drift (EDD) direction to the ion diamagnetic drift (IDD) direction to occur on the same flux surface. For example, Figure 7 shows the measured turbulence phase velocities in the poloidal direction with two different diagnostics: the GPI in red and the MLP in black. The data are from shot no. 1160616009. The measured quantities are shown as a function of the normalized radius or $\rho$ where $\rho = R - R_{LCFS}$ at the outer midplane. $R_{LCFS}$ is calculated from EFIT. Both diagnostics indicate that the direction of the measured phase velocity changes from the positive (EDD) to the negative (IDD) direction. Since this is an actual physical process, this radial location is assumed to occur at the same flux surface regardless of the measurement locations or diagnostic types. In the figure, this assumption is satisfied after shifting the GPI measurement by 0.8 cm. The same amount of the radial shift is consistently found to be needed in matching a number of GPI measurements with the MLP measurements. Therefore, when used as inputs to DEGAS 2 the $n_e$ and $T_e$ profiles are shifted radially by -0.8 cm. The MLP profile is shifted with respect to the GPI view location because the GPI views are fixed in the R and Z space, and independent of the plasma profile. We applied the default radial shift in the remaining three plasma discharges studied in this paper because in these shots the MLP was not plunged deep enough to observe the "flip". However, for the $D_2$ puff shot, we find that the radial shift of -0.4 cm best matches to the experimental measurement, as discussed in the next section.

## 4. Comparison of the Experimental and Simulation Brightness Results

### 4.1. Deuterium-puff

In the $D_2$ case puff case, good agreement is found between the experiment and simulation in terms of the peak brightness and spatial profile within the uncertainty of profile shift of ~ 0.4 cm. Figure 8(a) compares the $D_\alpha$ brightness profile along the major radial direction at the vertical location where the brightness is maximum. As discussed in Section 2.1, the intrinsic emission (dashed black line) inferred from the sister discharge is subtracted from the recorded $D_\alpha$ brightness (thin black line), in order to calculate the brightness profile due to puff only (thick black line). The modeled brightness profile with the default shift of -0.82 cm (blue line with a triangle marker), however, does not capture the experimental peak location at R = 89 cm. As shown in Figure 8(b), the low electron temperature limits the emission level. Because of the unknown radial shift required for this particular shot (the MLP was not deep enough to see the "flip"), a different radial shift is applied to the input kinetic profiles. The best agreement is found when the profile is shifted by -0.41 cm (red line with a circle marker). As shown in Figures 8(b) and (c), the density and temperature are slightly increased as compared to the default case. Figure 9 compares the $D_\alpha$ brightness profiles along the Z direction at R = 89.1 cm. The modeled profile captures the broadness observed in the profile along the vertical direction due to the use of the four capillaries in the nozzle below the midplane.

### 4.2. Helium-Puff

In the helium puff cases, the helium line emission predicted by DEAGS 2 is found to be higher by a factor three over the wide density range studied in this paper. The profile shape is not in good agreement either. Figures 10, 11 and 12 compare the simulated brightness profiles to the experimental measurements at three different line-averaged densities. The measured $n_e$ and $T_e$ profiles are shifted by -0.82 cm. To match the experimental brightness, it is found that the input temperatures need to be reduced by about half to suppress the emission. Note that the emission curve is a strong function of temperature below $T_e = 15$ eV.

Figure 10 compares the helium I line brightness profiles between the experiment and modeling at the lowest density (shot no. 1160616009). As shown in Figure 10(a), the predicted brightness peak is about three times higher than the experimental peak, and its radial location is shifted slightly outward by a half centimeter in the default case (blue line with a square marker). As shown in Figure 10(b), the rapid decrease in the predicted brightness at $R < 89$ cm is due to the strong ionization effect with $T_e$ above 20 eV, and the absolute level matches to the experimental result in this region. Thus, the focus was given to match the brightness level at $R > 89$ cm by reducing the input temperature by half, as shown by the profile in green in Figure 10 (b). This results in a good match in the brightness level at the outer view location (green line with a circle marker), while there remains a discrepancy at the inner view location by a factor of three. Raising the input density can also lower the emission rate, but its effect is found to be modest compared to that with the reduced temperature. Figure 10(c) shows the brightness profile along the Z direction at $R = 89.1$ cm. All three modeling results capture the broadness of the experimental profile, but the discrepancy in the peak brightness persists within the given variations in the input plasma profiles.

Figure 11 compares the helium I line brightness profiles at a medium density (shot no. 1160616016). As shown in Figure 11(a), the predicted profile with the default kinetic profile (in blue) does a poor job in matching the experimental result, except at the inner radial view location where the electron temperature is high enough so that ionization becomes important. The input temperature profile is modified by reducing the temperature where $T_e$ is less than 20 eV, as shown in Figure 11(b). This modification in temperature leads to a better match. The over-prediction in the brightness at $R < 89$ cm is due to the increase in neutral population with the reduced ionizations. Figure 11(c) shows that the reduction in temperature also leads to the better agreement in the brightness profile along the vertical direction.

Figure 12 compares the helium I line brightness profiles at the highest density (shot no. 1160616022). Figure 12(a) indicates that the experimental profile is a linearly decreasing function of the major radius, whereas the modeling predicts a structure in the profile with a higher peak brightness. When the input temperature is reduced by half (Figure 12(b)), this structure diminishes and the absolute brightness level also matches to the experimental result. Unlike the previous two cases, the emission at an inner view location is weak due to lower temperatures below 20 eV. Figure 12(c) shows that the brightness profile along the vertical direction at R = 89.1 cm matches the experimental profile with the reduced electron temperature.

5. **Discussion**

In the previous section, we compared the helium I line and $D_\alpha$ brightness measurements to the DEGAS 2 modeling results. While a good match is found for the $D_2$ puff case, the reduction in the input temperature was necessary to reasonably match the He I brightness profile to the experimental measurement. In this section, two possible causes for this discrepancy are discussed. First is local cooling of the plasma due to the interaction of electrons with neutrals. Second is the effect of turbulence on neutrals. In the SOL, time-dependent turbulence effects are known to affect neutral penetrations [16-18], whereas the stationary steady state plasma is assumed in the model studied in this paper.

The strength of local cooling is examined with the electron power loss calculated in the DEGAS 2 code. Note that the ionization is the only electron power loss channel in the modeling. Figure 13(a) shows the slice-cut of the volumetric electron loss rate in Watts/m$^3$ at Z=-0.03 m for the high density helium puff shot (shot no. 1160616022) in front of the nozzle. Note that the X-direction in the figure is nearly the major radial direction, and the Y-direction is nearly the toroidal direction. The peak loss rate is ~ 2.5 MW/m$^3$ at the inner view location where the electron temperature is high enough for significant ionizations. Figure 13(b) shows the line-integrated electron loss rate along the Y-direction in W/m$^2$. This integration path rather than the path along toroidal direction is justified because the gas puff cloud has a short arc length (~ 0.1 m) as compared to the total arc length of the plasma (~2×π ×R= 5.65 m for R=0.9 m). The peak value in this case is found to be ~150 kW/m$^2$. This integrated power flux can be considered as a power sink at a given field line.

This heat loss is found to be comparable or larger than the parallel heat flux required to maintain the electron temperature in the low temperature region below 10 eV. The assumption here is that the cross-field driven heat flux is balanced by the parallel heat conduction along the field line in the SOL, which determines the upstream electron temperature ($T_u$) in the open-field region: $q_\parallel \approx k_0 T_u^{\frac{7}{2}}/(3.5\ L)$ where $k_0$= 2000 [W/m/eV$^{7/2}$] and L ≈ 10 m is the typical C-Mod connection length. For example, $q_\parallel \approx$ 200 kW/m$^2$ for $T_u$ = 10 eV. From the measured electron temperature shown in Figure 13(c), the required parallel heat flux to sustain this temperature is plotted in red in Figure 13(b). In the low temperature region below ~ 10 eV, the heat loss due to gas puff is larger than the input heat flux from the plasma. This indicates that gas puff can locally cool the plasma within the field lines intercepting the gas cloud, and could justify the reduction in the input temperature applied in the previous section. Note that the discussion in [24] focuses on the temperature perturbation in the confined plasma within the LCFS. Note that the effect of gas puff on the density perturbation is not expected to be large. Assuming that all the neutrals are ionized (S~10$^{20}$ #/s) within the gas cloud (L = 0.06 m along the toroidal direction, and gas cloud volume V= L×X×Y ≈ 0.06×0.03×0.1 = 1.8×10$^{-4}$ m$^3$), and that the newly created electrons stream along the field-line out of the gas cloud at the plasma sound speed $c_s \approx \sqrt{\frac{2T_e}{m_i}} \approx 2.2 \times 10^4\ m/s$ for $T_e$ = 10 eV and He$^+$ ions, one finds that δne = (S×L/V)/ (2 $c_s$)= 7.6×10$^{17}$ m$^{-3}$ << 1×10$^{19}$ m$^{-3}$.

On the other hand, one issue with this interpretation is the absence of such effects with the $D_2$ puff. Modeling results indicate that the decrease in the input electron temperature by half for the $D_2$ case suppresses the emission significantly at the outer view region, and as a result the agreement with the experimental profile becomes worse. A similar analysis on the electron power loss also indicates that for $T_e < 10$ eV the electron power loss dominates over the required parallel heat for the given temperature. Unfortunately, the MLP measurement location is not magnetically connected to the gas puff region, so the local temperature at the field line intercepting the gas cloud is not available in this study.

Several theoretical and modeling studies [16-18] show that plasma turbulence can modify the neutral particle spatial distribution when the neutral mean free path is on the order of or less than the turbulence correlation length. The SOL density turbulence with large fluctuation levels above 50% is shown to affect the time average ionization source term, and the neutral density decays slowly in the presence of turbulence, indicating better neutral penetration. On the other hand, it is also reported that the fluctuations can increase the ionization rate below the ionization threshold temperature because the density and temperature fluctuations in the SOL are usually in-phase with blobby like transport [17]. This is examined with the available MLP measurements in a dwell mode for a series of stationary C-Mod plasmas at different Greenwald fractions. The MLP was about ~ 1 cm away from the EFIT LCFS. The relative helium ionization fluctuation (R) is found to increase from 0.5 to 1.6 as the Greenwald fraction increases from 0.14 to 0.7. The effective helium ionization rate increases with the Greenwald fraction because of the in-phase nature of the density and temperature fluctuations. Note that the ionization rate increases with the density and temperature. On the other hand, the emission rate is known to decrease with density. Further, the helium emissivity is peaked at ~20 eV. As a result, blobby fluctuations around at 20 eV (i.e., typical SOL temperatures) effectively lower the emissivity. For example, at the Greenwald fraction of 0.7, the time-averaged ionization can be ~40% higher than the ionization rate found from the averaged $n_e$ and $T_e$, while the time-averaged emission rate can be ~10% lower than that found from the averaged $n_e$ and $T_e$. Assuming these effective increase in the ionization rate and the decrease in the emission rate coefficient in a simple 1D model, the effective emission is found to decrease by ~ 25 %. These functional dependences of the ionization and emission rates are in the correct direction to explain the discrepancy between the model and the experiment.

To further examine the difference between the helium and $D_2$ puff cases, the Kubo number [17] is examined. The Kubo number K quantifies the strength of time-dependent turbulence effects on the neutral propagation: $K = R \times a$ where $R=\sigma^{1/2}/<v>$ is the ratio of the standard deviation in the ionization rate ($\sigma^{1/2}$) to the time-averaged ionization rate ($<v>$), and $a=\lambda/L$ is the ratio of the turbulence correlation length ($\lambda$) to the neutral mean free path ($L = v_0/<v>$). Here $v_0$ is the neutral speed. It is generally considered that effects of fluctuations becomes important for $K >> 1$. A clear difference between the helium and deuterium atom is found in the latter term ($a= \lambda/L$). A deuterium atom has a higher energy (~ 3 eV) associated with the Frank-Condon dissociation and charge-exchange processes, and thus has a longer mean free path than the helium atom. In contrast, helium atoms gain little energy from their interactions with the background ions.

Figure 14 compares the ionization mean free paths of a neutral helium atom at 0.026 eV and a deuterium atom at 3 eV as a function of the electron temperature and density. The reaction rates are from the database used by DEGAS 2. At $T_e = 25$ eV and $n_e = 3 \times 10^{19}$ m$^{-3}$, the ionization mean free path of the deuterium atom is ~ 3 cm as compared to ~ 0.5 cm of the helium atom. Assuming that all other parameters in the Kubo number are the same, with λ = 0.5 ~ 1 cm and R ≈ 1, the short mean free path of the helium atom results in K ≳ 1, which can be larger than that of the deuterium atom by a factor of six. The Kubo number for the 3 eV deuterium atom is below 1 for typical C-Mod SOL plasma parameters because its mean free path is greater than 1 cm. This difference in the Kubo number is in line with the better agreement seen in the $D_2$ puff case between the experiment and steady-state simulation, and suggests that time-dependent turbulence effects might need to be considered for the helium puff case.

Due to the cessation of the C-Mod operation, the observed discrepancy cannot be further investigated experimentally on C-Mod. In retrospect, other helium line emissions could have been measured and compared with the modelling results for the completeness. In this way, it could have been possible to identify whether the observed discrepancy is line-specific or not. In addition, $D_\alpha$ brightness measurements at higher Greenwald fractions would have helped understand the role of turbulence on neutral penetrations. Furthermore, it would be worthwhile directly measuring the local electron temperature and density in the neutral gas cloud.

## 6. Summary

In this paper, the $D_\alpha$ and helium line emission profiles measured in the C-Mod GPI experiments are compared with the synthetic emission profiles from a neutral transport code DEGAS 2. The GPI experimental setup on C-Mod provides an ideal condition to study neutral transport of deuterium and helium neutrals. The experimental calibration processes are detailed in order to compare the absolute level of the line emissions. The injected flow rate is constructed by combining the waveforms from the GPI signal and the in-vessel pressure measurement. Individual detectors are calibrated using a continuum light source. The electron density and temperature profiles are measured with the Thomson and scanning probe. The measured $D_\alpha$ peak brightness and profile shape agree well with the simulation results. However, the modelled helium I line brightness is higher by a factor of three than the experimentally measured one over the wide range of plasma density. The discrepancy is large where the electron temperature is low. The neutral-ion scattering process itself does not reduce the neutral density enough to match the measured helium line brightness. Local cooling due to the injected gas puff could have lowered the electron temperature and suppress the emission, as indicated by the comparison of the input heat flux with the electron power loss rate. Another possible cause is the effect of turbulence that is not modeled in this study. The cold helium atom is susceptible to the turbulence effect due to its short ionization mean free path. Further experimental and modeling investigations are needed to quantitatively understand this discrepancy.

**Acknowledgements**

This research was conducted on Alcator C-Mod, which is a DOE Office of Science User Facility, and is supported in part by: US Department of Energy, contracts DE-FC02-99ER54512 and DE-AC02-09CH11466.

Table 1: A summary of the four plasma discharges studied in this paper. The central magnetic field is 5.4 T.

| Shot No. | Puff Gas Type | $\bar{n}_e$ ($10^{20}$ m$^{-3}$) | $I_p$ (MA) | $\bar{n}_e/n_G$ |
|---|---|---|---|---|
| 1160617031 | $D_2$ | 1.5 | 1.2 | 0.19 |
| 1160616009 | He | 0.8 | 0.5 | 0.24 |
| 1160616016 | He | 1.6 | 0.5 | 0.49 |
| 1160616022 | He | 2.3 | 0.5 | 0.70 |

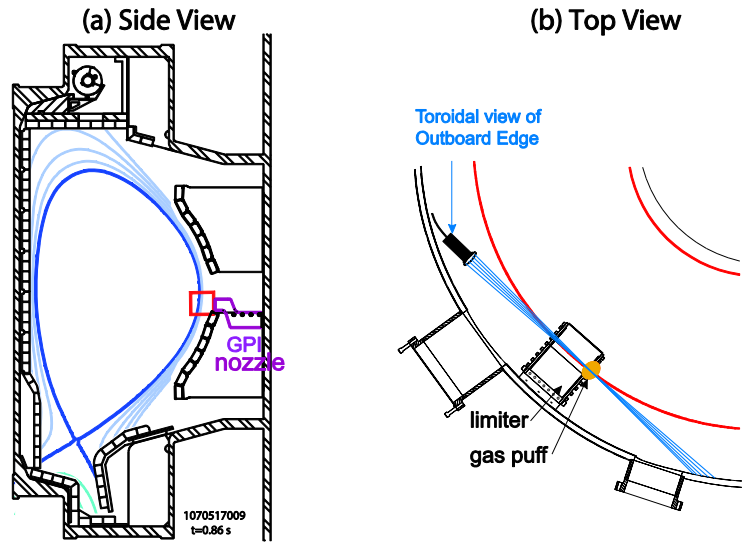

Figure 1: (a) Side view of the Alcator C-Mod tokamak that shows the GPI nozzle location at the outer midplane, (b) Top view of the Alcator C-Mod tokamak that shows the front-end optics system and the gas puff nozzle next to the limiter.

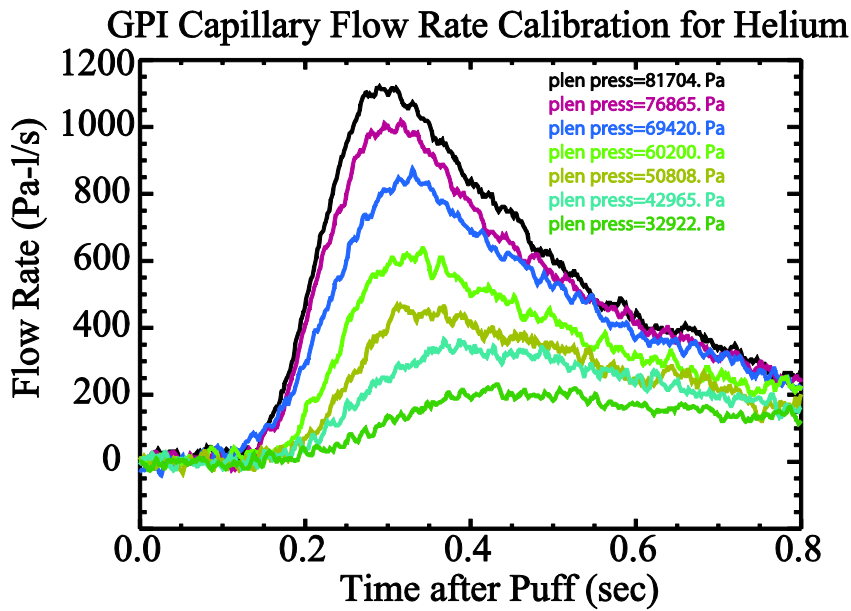

Figure 2: GPI capillary nozzle flow rate measurements for different NINJA plenum pressures derived from in-vessel pressure gauge measurements located well away from the GPI nozzle..

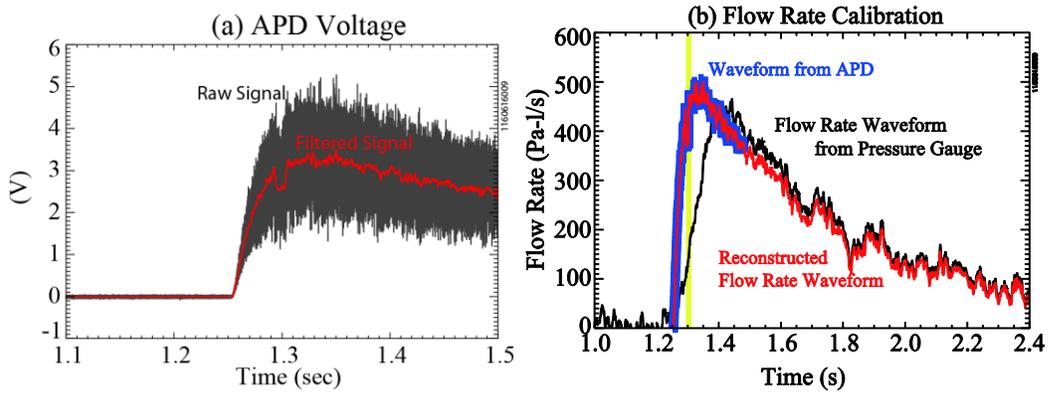

Figure 3: (a) the APD voltage trace as a function of time. Black: the measured APD signal (helium I line emission) in a deuterium plasma (shot no. 1160616009). Red: the filtered measured APD signal. (b) Reconstructed flow rate waveform in red for the He puff into the plasma with the lowest density (shot no. 1160616019). The waveform in blue is proportional to the He I emission for a view in front of the nozzle for that discharge, while the waveform in black is the time rate of change of pressure from a re-enactment of the gas puff into the C-Mod vacuum chamber.

| Shot | Gas | Plenum Pressure (Torr) | Trigger time (sec) | Puff Duration (sec) | Total injected particles (Pa-l) | Flow rate at 1.3 sec (Pa-l/s) | Flow rate at 1.3 sec ($10^{20}$ #/sec) |
|---|---|---|---|---|---|---|---|
| 1160616009 | He | 385 | 1.10 | 0.09 | 296 | 467 | 1.11 |
| 1160616016 | He | 381 | 0.96 | 0.09 | 295 | 367 | 0.87 |
| 1160616022 | He | 375 | 0.96 | 0.09 | 292 | 373 | 0.90 |
| 1160617031 | $D_2$ | 392 | 1.11 | 0.08 | 311 | 613 | 1.49 |

Table 2: A summary of the neutral gas injections for the four plasma discharges studied. The density and temperature profiles are measured at 1.3 second.

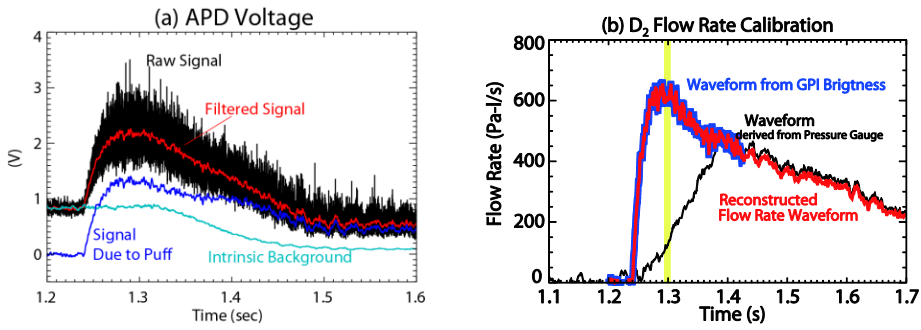

Figure 4: (a) APD voltage trace as a function of time. Black: the measured APD signal ($D_\alpha$ emission) in a deuterium plasma (shot no. 1160617031). Red: the filtered measured APDI signal. Cyan: the filtered APD signal from another discharge without $D_2$-puff (shot no 1160617030). Blue: the corrected APD signal after subtracting the background contribution. (b) Flow rate waveform in red for the $D_2$ puff into the plasma. The waveform in blue is proportional to the $D_\alpha$ emission (the blue trace in (a)) due only to puff for a view in front of the nozzle for that discharge, while the waveform in black is the time rate of change of pressure from a re-enactment of the gas puff into the C-Mod vacuum chamber.

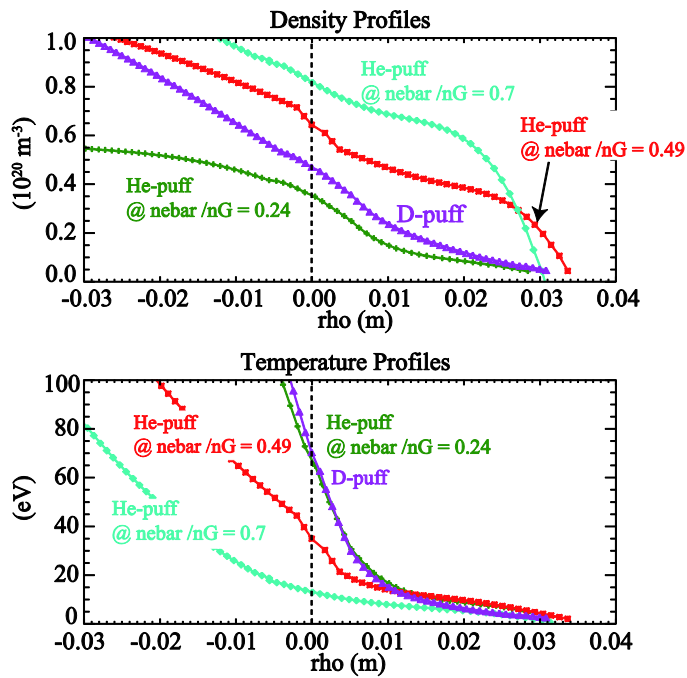

Figure 5: Density (Top) and temperature (bottom) profiles in the edge/SOL region measured with the Thomson and MLP for the four plasma discharges studied in this paper.

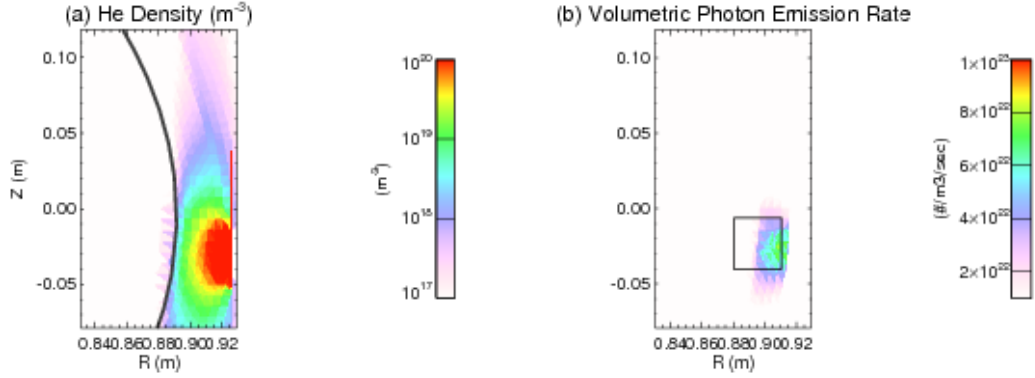

Figure 6: (a) Side-cut of the 2D distribution of neutral helium density and (b) side-cut of the 2D distribution of volumetric photon emission rate in the R-Z space in front of the nozzle from DEGAS 2 for the shot no. 1160616022. The black curve in (a) is the plasma EFIT separatrix, and the black rectangle in (b) is the field-of-view of the detector at the target plane.

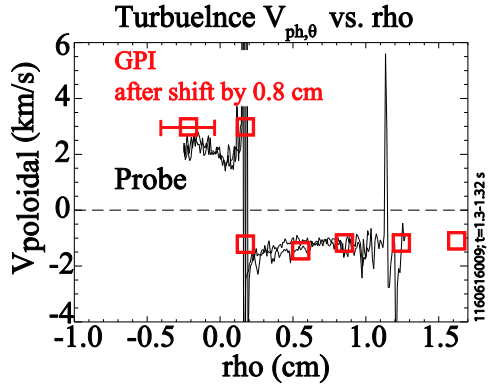

Figure 7: The turbulence poloidal phase velocity measurement with the GPI in red and the mirror Langmuir probe in black. Both measurements are mapped to the outer midplane. To align the "flip" observed in the direction of the measured poloidal phase velocity, the GPI measurement is radially shifted. Here, $\rho = R - R_{LCFS}$.

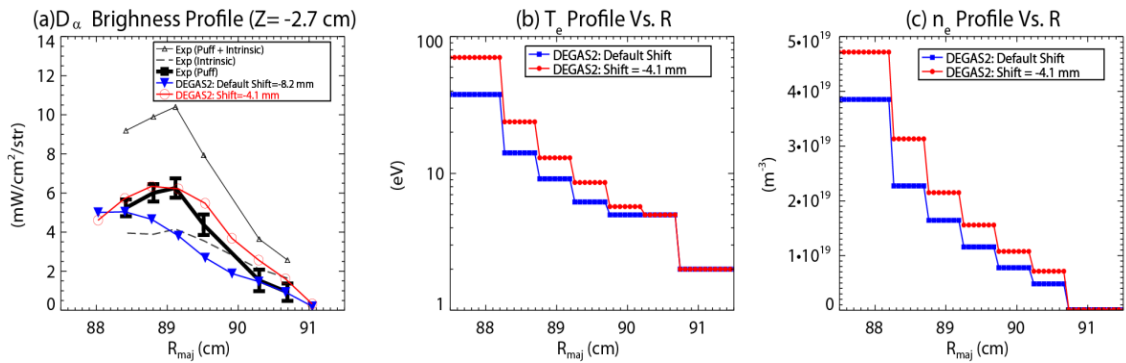

Figure 8: (a) A comparison of the experimentally measured $D_\alpha$ brightness profile as a function of the major radial direction at Z = -2.7 cm in thick black with the synthetic diagnostic result from DEGAS 2 in blue with the radial shift of the input profile by -0.82 cm and in red with the radial shift of the input profile by -0.41 cm. The profile for the intrinsic $D_\alpha$ brightness is from a sister shot without $D_2$ puff. The profile for $D_\alpha$ brightness due to puff is determined by subtracting the intrinsic $D_\alpha$ brightness from the measured $D_\alpha$ brightness. (b) The input electron temperature profiles in blue with the radial shift of -0.82 cm and in red with the radial shift of -0.41 cm. (c) The input electron density profiles in blue with the radial shift of -0.82 cm and in red with the radial shift of -0.41 cm.

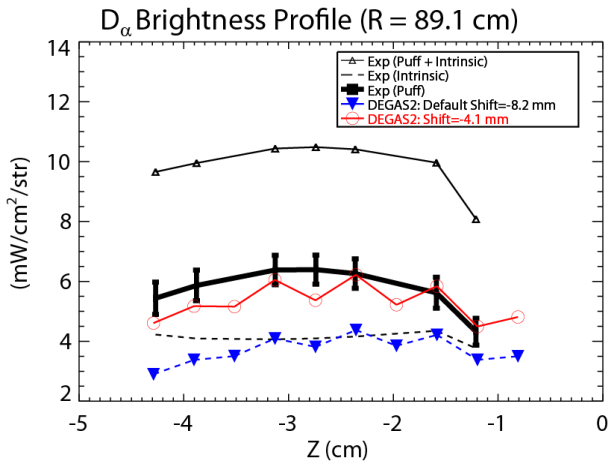

Figure 9: A comparison of the experimentally measured $D_\alpha$ brightness profile as a function of the vertical direction at R = 89.1 cm in thick black with the synthetic diagnostic result from DEGAS 2 in blue with the radial shift of the input profile by -0.82 cm and in red with the radial shift of the input profile by -0.41 cm. The profile for the intrinsic $D_\alpha$ brightness is from a sister shot without $D_2$ puff. The profile for $D_\alpha$ brightness due to puff is determined by subtracting the intrinsic $D_\alpha$ brightness from the measured $D_\alpha$ brightness.

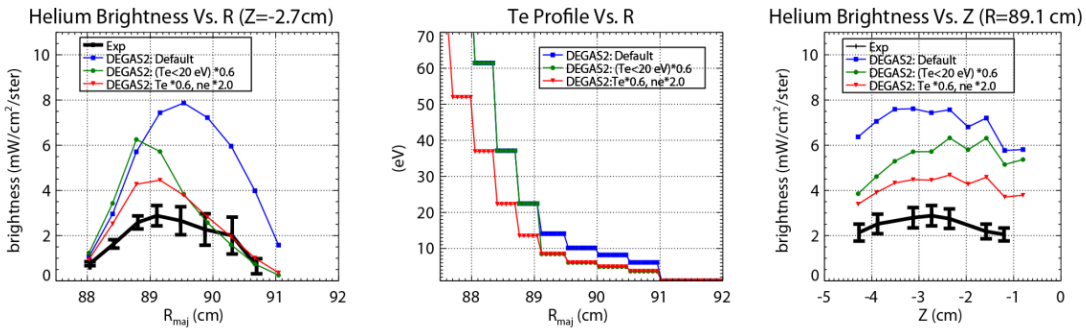

**Figure 10:** (a) A comparison of the experimentally measured He I line brightness profile for the shot no. 1160616009 as a function of the major radial direction at Z = -2.7 cm in thick black with the synthetic diagnostic results from DEGAS 2 in blue with the default input profiles, in green

with the electron temperature reduced by 40% for $T_e < 20$ eV, and in red with the reduction in the electron temperature by 40% and the doubling the electron density. (b) The electron temperature profiles in blue with the default input, in green with the electron temperature reduced by 40% for $T_e < 20$ eV, and in in red with the reduction in the electron temperature by 40% and the doubling the electron density. (c) A comparison of the experimentally measured He I line brightness profile for the shot no. 1160616009 as a function of the vertical direction at R = 89.1 cm in thick black with the synthetic diagnostic results from DEGAS 2 in blue with the default input profiles, in green with the electron temperature reduced by 40% for $T_e < 20$ eV, and in red with the reduction in the electron temperature by 40% and the doubling the electron density.

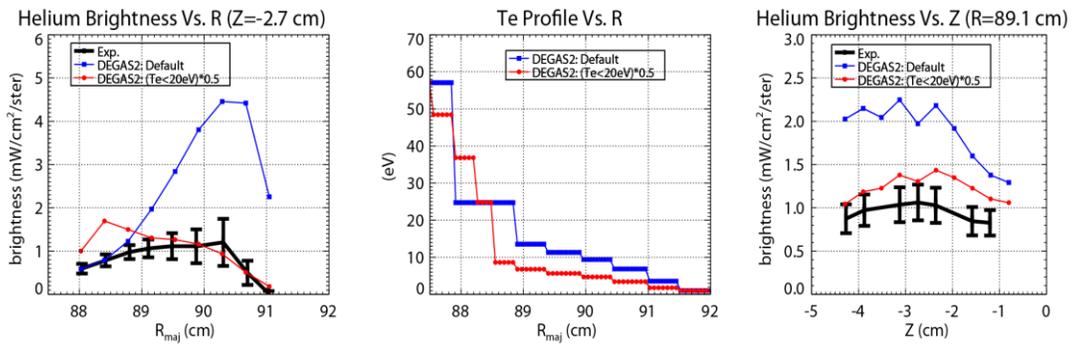

**Figure 11:** (a) A comparison of the experimentally measured He I line brightness profile for the shot no. 1160616016 as a function of the major radial direction at Z = -2.7 cm in thick black with the synthetic diagnostic results from DEGAS 2 in blue with the default input profiles, and in red with the reduction in the electron temperature by 50% for $T_e < 20$ eV. (b) The electron temperature profiles in blue with the default input, and in red with the electron temperature reduced by 50% for $T_e < 20$ eV. (c) A comparison of the experimentally measured He I line brightness profile for the shot no. 1160616016 as a function of the vertical direction at R = 89.1 cm in thick black with the synthetic diagnostic results from DEGAS 2 in blue with the default input profiles, and in red with the electron temperature reduced by 50% for $T_e < 20$ eV.

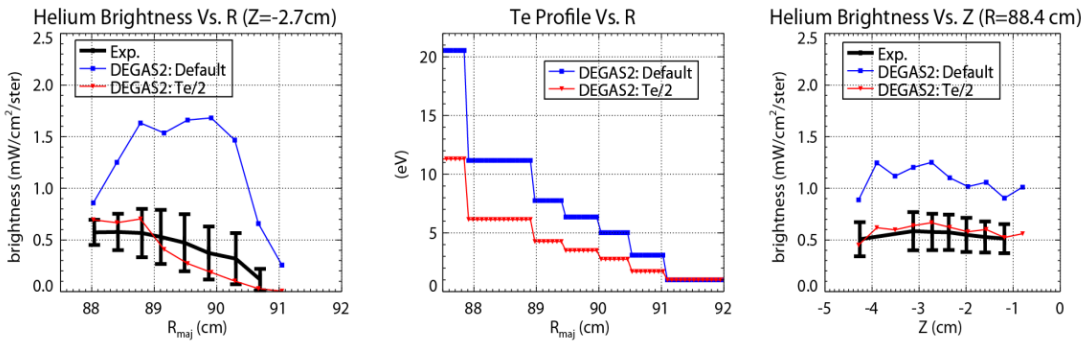

**Figure 12:** (a) A comparison of the experimentally measured He I line brightness profile for the shot no. 1160616022 as a function of the major radial direction at Z = -2.7 cm in thick black with the synthetic diagnostic results from DEGAS 2 in blue with the default input profiles, and in red with the reduction in the electron temperature by 50%. (b) The electron temperature profiles in blue with the default input, and in red with the electron temperature reduced by 50%. (c) A comparison of the experimentally measured He I line brightness profile for the shot no. 1160616022 as a function of the vertical direction at R = 89.1 cm in thick black with the synthetic diagnostic results from DEGAS 2 in blue with the default input profiles, and in red with the electron temperature reduced by 50%.

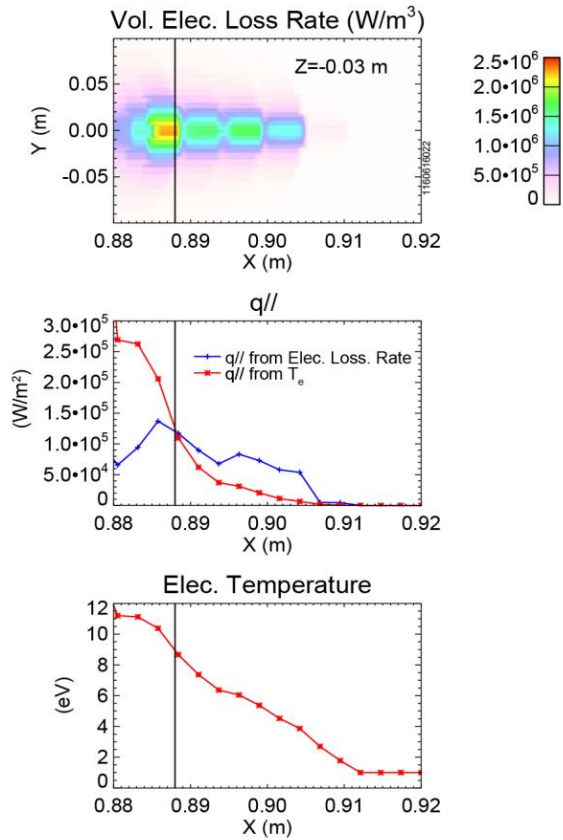

**Figure 13:** (a) The volumetric electron loss rate (W/m$^3$) as a function of X and Y in the DEGAS 2 simulation grid at Z = -3 cm. The X-direction is nearly the major radial direction, and the Y-direction is nearly the toroidal direction. (b) The line-integrated heat flux along the Y-direction of the electron power loss rate as a function of X in blue, and the parallel heat flux to maintain the measured electron temperature in red. (c) The line-averaged electron temperature along the Y-direction as a function of X.

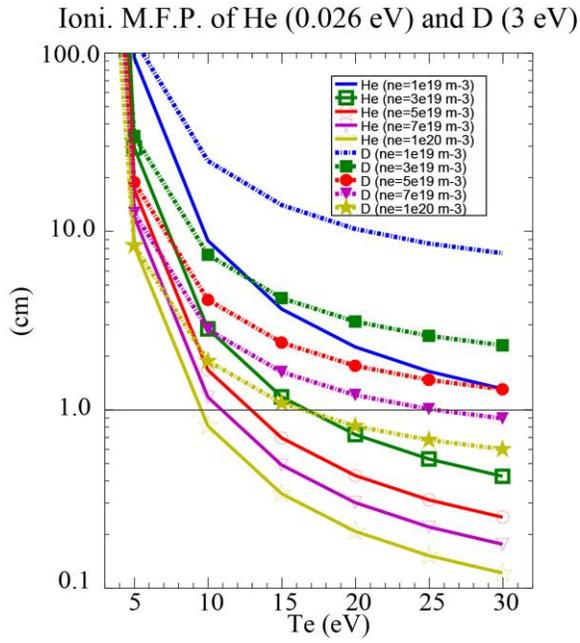

**Figure 14:** Ionization mean free paths of the 0.026 eV helium atom (He) in solid lines and the 3 eV deuterium atom (D) in dashed lines as a function of the electron temperature at four different electron densities: 1 (blue), 3 (green), 5 (red), 7 (purple), and 10 (yellow)$\times 10^{19}$ m$^{-3}$.